\documentclass[twocolumn,showpacs,preprintnumbers,amsmath,amssymb]{revtex4}
\usepackage{graphicx}
\usepackage{dcolumn} 
\usepackage{bm} 
\usepackage{color}

\begin{document}

\title{Experimental Determination of Electron Transition Probabilities in Elementary Electron-Phonon Scattering Processes Using Electron-Energy-Loss Spectroscopy: \\The Example of Graphite
} 

\author{Hirofumi Yanagisawa${}^{1,2}$}
\email{hirofumi@physik.uzh.ch}
\author{Toma Yorisaki${}^{2}$}
\author{Ryota Niikura${}^{2}$}
\author{\\Shuhei Kato${}^{2}$}
\author{Yasuchika Ishida${}^{2}$}
\author{Kenji Kamide${}^{3}$}
\author{Chuhei Oshima${}^{2}$}
\affiliation{${}^1$Physik Institut, Universit\"{a}t Z\"{u}rich, Winterthurerstrasse 190, CH-8057 Z\"{u}rich, Switzerland \\
${}^2$\mbox{Department of Applied Physics, Waseda University, 3-4-1 Okubo, Shinjuku-ku Tokyo 169-8555, Japan} \\
${}^3$Department of Physics, Osaka University, Toyonaka, Osaka 560-0043, Japan
}

\date{\today}

\begin{abstract}
We have investigated coupling constants in elementary electron-phonon scattering processes on a graphite surface by the combined use of high-resolution electron-energy-loss spectroscopy (HREELS) and very low-energy electron diffraction (VLEED). HREELS is used to measure the modulations of electron transition probabilities from incoming electrons in vacuum to outgoing electrons in vacuum where the transition includes one-phonon scattering processes inside a solid. Determining the electronic band structures of graphite with VLEED, we defined electronic states of the solid surface that electrons entered before and after scattering off phonons. Thus, we observed that the measured electron transition probabilities significantly depended on whether the electrons were in a bulk Bloch state or an evanescent state before scattering off the phonons. This result clearly indicates that the measured electron transition probabilities reflect the strength of the coupling constants in the solid. 
\end{abstract}

\maketitle

The coupling constants of elementary scattering processes in solids are the most fundamental parameters in condensed matter physics. Electron-phonon coupling constants determine, for instance, critical temperatures in BCS superconductors~\cite{bcs:1,review1,zimantext,epi1:1} and a time scale for energy dissipation in the hot-carrier dynamics of metals and semiconductors~\cite{haugtext,lisowski04,carbone09}. Nevertheless, it is quite difficult to detect the elementary scattering processes in solids because they are many-body systems. To date, experimentally obtained electron-phonon coupling strength have been integrated over wave vectors and the energies of both electron and phonon systems~\cite{epi1:1}, or either of them~\cite{epi:2, epi:3}. Therefore, the detailed microscopic features of the elementary processes remain unknown. Here, we present the first experiments directly addressing elementary electron-phonon scattering processes. 

\begin{figure}[b]
\vspace{-5pt}
\includegraphics[scale=0.28]{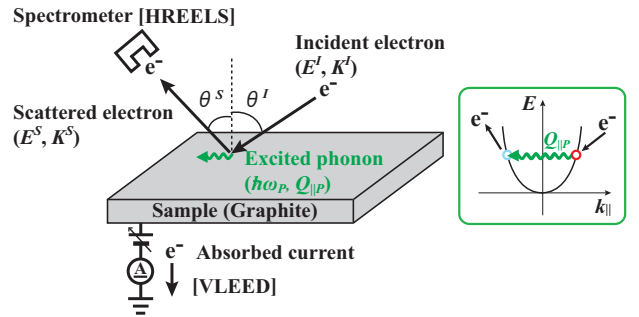}
\vspace{-5pt}
\caption{\label{fig:epsart}
(color online) Schematic diagram of experiments. In HREELS, an incident electron excites a phonon at the surface, and the loss energies are measured by the analyzer. In VLEED, a picoammeter is used for measuring the sample current modulation as a function of $E^{I}$ and $K^{I}_{\parallel}$. The inset framed by a rectangle shows a conceptual diagram of the electron scattering process in HREELS experiments. The parabolic curve indicates a electric band, and the red and blue circles indicate the electronic states that electrons excite in the solids before and after scattering off phonons, respectively.}

\end{figure}

To address the elementary processes in such detail, we employed HREELS. It is widely used for measuring surface-phonon energies via electron scattering experiments~\cite{hreelsbook:1,hreels:1, hreels:2}, and the scattering geometry for a one-phonon-excitation process is shown in Fig.~1. The coupling constants can be extracted from the intensities of scattered electrons at a particular solid angle because the scattering process involves an electron-phonon scattering process inside the solid; the incident electrons that penetrate into the solid excite the electronic states of the solid surface, as illustrated in the inset of Fig. 1~\cite{phononintensity:1, tong75}. By the scattering of phonons at a rate determined by the electron-phonon coupling constant, electrons are transferred to other electronic states before bouncing from the surface to the vacuum~\cite{phononintensity:1, hufner95}. Therefore, the intensities of outgoing electrons should depend on the strength of the electron-phonon coupling in the solid. The specific electronic states excited in the solid are defined by wave vectors parallel to the surface, $K^{I}_{\parallel}$ and $K^{S}_{\parallel}$, as well as the energies $E^I$ and $E^S$ of the incoming and outgoing electrons in the vacuum, respectively. The conservation laws hold for electrons traveling between the vacuum and the solid; the energies and wave vectors of scattered phonons $(\hbar \omega_{P},Q_{\parallel P})$ can be obtained from the conservation laws of the scattering~\cite{hreelsbook:1}:
\begin{eqnarray}
\hbar \omega_{P}=E^{I} - E^{S}, \\
Q_{\parallel P}=K^{I}_{\parallel} - K^{S}_{\parallel}.
\end{eqnarray}
If the electronic structures are known, the detailed features of the electron-phonon coupling should be addressable. VLEED can be used to measure the electronic energy bands in solids by measuring the sample current modulation with an ammeter, as in Fig. 1~\cite{vleed:1,vleed:2}.

Using HREELS, we investigated electron-phonon coupling constants by measuring the electron transition probabilities from incoming electrons in vacuum to outgoing electrons in vacuum; the transition processes involved one-phonon scattering processes at a graphite surface. The electronic structures of graphite were determined by VLEED. From these data sets, we examined the dependence of the electron transition probabilities on the electronic states of graphite. We observed a significant dependence on the state of the electrons before the phonon scattering occurred. This result clearly indicates that the measured electron transition probabilities reflect the strength of the coupling constants in the solid.

The electron transition probabilities via one-phonon scattering processes in HREELS experiments were determined from the intensities of outgoing electrons using the formula of Li {\it et al.}~\cite{phononintensity:1}. Let $\left[ dS_{P} \left( K^{I}, K^{S} \right) / d \Omega \right] d \Omega$ be the {\it fraction} of incoming electrons that emerge into the solid angle $d \Omega$ after scattering inelastically off the vibrational mode of polarization P. According to Ref.~\cite{phononintensity:1},
\begin{eqnarray}
\frac{d S_{P} \left( K^{I}, K^{S} \right)}{d \Omega} \propto &&  \frac{E^{I} \cos^{2} \theta_{S}}{\cos \theta_{I}} \left( n_{P} + 1 \right) \left( \frac{1}{\omega_{P}} \right) \nonumber \\ 
&& \times \left| \frac{\partial f}{\partial R_{\omega_{P} Q_{\parallel P}}} \right| ^{2},
\end{eqnarray}
where $n_{P} = \left[ \exp \left( \hbar \omega_{P} / k_{B} T \right) -1 \right] ^{-1}$ is the Bose-Einstein factor and $\left| \partial f / \partial R_{\omega_{P} Q_{\parallel P}} \right|$ gives the electron-phonon scattering amplitude in HREELS experiments, namely, the scattering amplitude from the initial state of incoming electrons in vacuum to the final state of outgoing electrons in vacuum when a crystal potential is deformed by lattice displacements $R_{\omega_{P} Q_{\parallel P}}$ because of a phonon mode of a unit amplitude with phonon energy $\omega_{P}$ and wave vector $Q_{\parallel P}$. We can say that the square of $\left| \partial f / \partial R_{\omega_{P} Q_{\parallel P}} \right|$ is the electron transition probability in the one-phonon-excitation process of HREELS experiments. To investigate the coupling constants in solids, we evaluated the relative values of the electron transition probabilities extracted from Eq.~(3).

The experiments were carried out in an ultrahigh vacuum chamber (base pressure: $1 \times 10^{-8}$ Pa) equipped with HREELS (Specs delta0.5). A single crystal of graphite (grown on an SiC(0001) substrate) was employed as a sample because the electronic and phonon structures of single-crystal graphite have been well studied~\cite{graphitefabrication:1,graphitephonon:1,hreels:2,vleed:2,vleed:4}. A picoammeter (ADVANTEST: 8252) was used for the VLEED measurements.

In the VLEED experiments, the sample current $I$ was measured while changing the incident electron energies $E^{I}$ for each angle (from 0 degrees to 75 degrees in 1-degree steps) along the $\overline{\Gamma}$-$\overline{\text{K}}$ direction in the two-dimensional surface Brillouin zone, depicted in the inset of Fig.~3(b). The sample current is modulated by the electronic structures. The simplest example of the modulation is a band gap. Because there are no bulk states available for incident electrons, the electrons are totally reflected, causing the sample current to exhibit a valley. In this way, modulation of the sample current $I(E^{I})$ exhibits the defining characteristic of electronic dispersion $E(k)$, and the critical points of $E(k)$ can be traced from the derivatives of $I(E^{I})$~\cite{vleed:1,vleed:2}. In Fig.~2(a), the VLEED spectra show a clear modulation of $dI/dE^{I}$. Electronic states exist between the local maxima and minima indicated by black and white triangles. 

\begin{figure}[b]
\vspace{-10pt}
\includegraphics[scale=0.36]{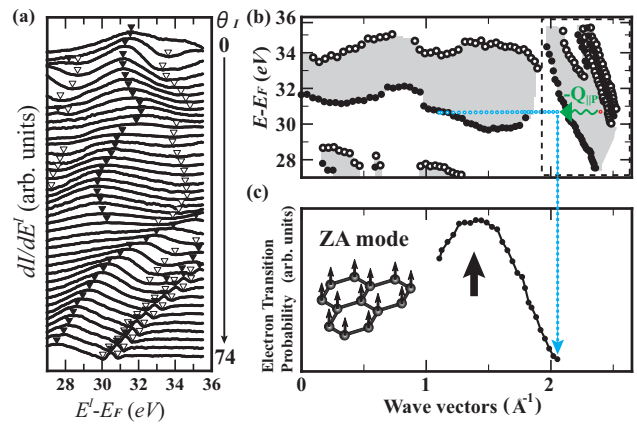}
\vspace{-10pt}
\caption{\label{fig:epsart}
(color online) (a) VLEED spectra of graphite for various incident angles measured along the $\overline{\Gamma}$-$\overline{\text{K}}$ direction in the two-dimensional surface Brillouin zone. The horizontal axis represents electron energies measured from the Fermi level, $E_{F}$. The black and white triangles indicate local maxima and minima. (b) An unoccupied band map of graphite determined by the VLEED spectra of (a). The wave vectors were determined by $K_{\parallel} = \left( 2mE^{I} \right)^{1/2}\sin \theta^{I} / \hbar$ . The black and white circles are the local maxima and minima of the spectra, and the shaded areas between them represent the existence of electronic states. (c) An example of an electron-transition-probability spectrum (see the text). }
\end{figure}

Figure 2(b) shows an unoccupied band map of graphite deduced from the VLEED spectra. Basically, this band map is a surface-projected dispersion $E \left( k_{\parallel} \right)$. Our VLEED data successfully reproduced the previous data~\cite{vleed:2}; we have, however, measured much deeper into the Brillouin zone. We observed a surface resonance state induced by a (11) diffraction beam parallel to the surface, as observed in Ref.~\cite{vleed:3} and represented by the rightmost, narrow band in the band map.

\begin{figure}[t]
\includegraphics[scale=0.32]{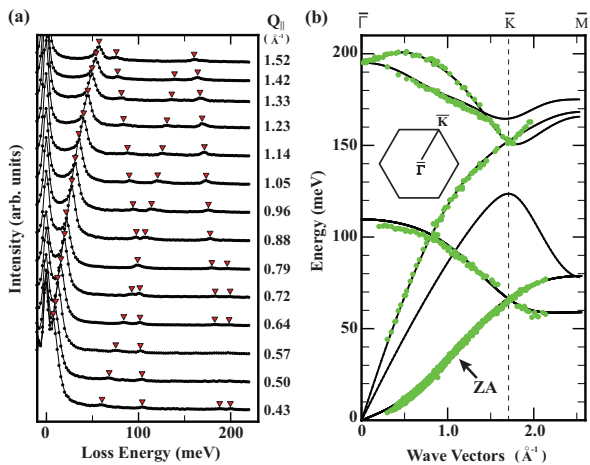}
\vspace{-10pt}
\caption{\label{fig:epsart}
(color online) (a)HREEL spectra of graphite for various wave vectors measured along the $\overline{\Gamma}$-$\overline{\text{K}}$ direction in the two-dimensional surface Brillouin zone, depicted in the inset of (b). The incident energy $E^{I}$ and the angle $\theta^{I}$ of the electron beam were 26.3 $eV$ and 72$^{\circ}$, respectively. The red triangles denote energy loss peaks. (b) Phonon energy dispersion curves of graphite. The experimental data (green circles) are compared with an {\it ab initio} calculation (solid lines).
}
\vspace{-10pt}
\end{figure}


Figure 3(a) shows typical HREEL spectra of graphite for various wave vectors along the $\overline{\Gamma}$-$\overline{\text{K}}$ direction. As denoted by the triangles, clear loss features appear in the spectra. The intensities of the loss peaks are proportional to electron transition probabilities, as described in Eq.~(3). From these spectra, we determined the phonon dispersion relation experimentally. In Fig. 3(b), the measured phonon dispersion curves (green dots) are compared with theoretical ones derived from {\it ab initio} calculations (solid lines) \cite{dacapo}. The details of the calculations are described elsewhere \cite{hreels:1, hreels:2}. The observed curves are in fairly good agreement with theory. These data ensure the reliability of our experimental techniques and justify the exclusion of additional scattering processes in graphite beyond one-phonon scattering.


In the constructed VLEED map, the initial and final states in the HREELS experiments can be plotted, representing incoming and outgoing electrons in vacuum. The initial-state wave vector and the energy are denoted by $K^{I}_{\parallel}$ and $E^{I}+\Phi$, respectively, where $\Phi$ is the work function of the sample and is equal to 4.75 $eV$ for graphite(0001). An example of an initial state is shown by a red circle in the band map in Fig.~2(b). The final states can be plotted on the band map using Eqs.~(1) and (2). Examples of the final states used in this work are plotted with blue circles. Acoustic phonons with out-of-plane polarization were employed to transfer electrons from the initial to final states and denoted by ZA in Fig.~3(b) \cite{hreels:2}; the intensity of electron scattering due to ZA phonons was the highest among all phonons under the given conditions of the incident electrons. 

At each final state, the relative values of the electron transition probability can be determined from the measured intensities of the outgoing electrons using Eq.~(3). Fig.~2(c) shows an electron-transition-probability spectrum, and the peak of the spectrum indicated by the black arrow means that {\it the initial incoming electrons optimally couple with the final outgoing electrons of this wave vector through one-phonon scattering processes.}  In addition, because the absolute values of outgoing electron intensities have to be measured~\cite{intensitymesurement:1}, we did not change any of the potential parameters of the spectrometer while measuring the single electron-transition-probability spectrum. The measurable spectral region in the final-state wave vectors was restricted by the rotation angle of the analyzer, $2 \theta^{I}+ \theta^{S} \leq 90^{\circ}$. The measured spectra were highly reproducible.

Fig. 4(a)-(c) shows the electron-transition-probability spectra taken from the three sets of initial states (a)-(c) shown by red circles in the magified VLEED band map of Fig.~4(d), in which the energy of each point in one set is the same. These initial states of incoming electrons excite the electronic states of graphite at the same points on the band map when they enter the solids. In the electronic band structures of graphite, white areas of the band map are gaps; wave functions here must be evanescent. On the other hand, the shaded area indicates a bulk Bloch state. Therefore, the electronic states of the white and shaded areas must have wave functions with different symmetries in the direction perpendicular to the sample surface. 

\begin{figure}[b!]
\vspace{-10pt}
\includegraphics[scale=0.42]{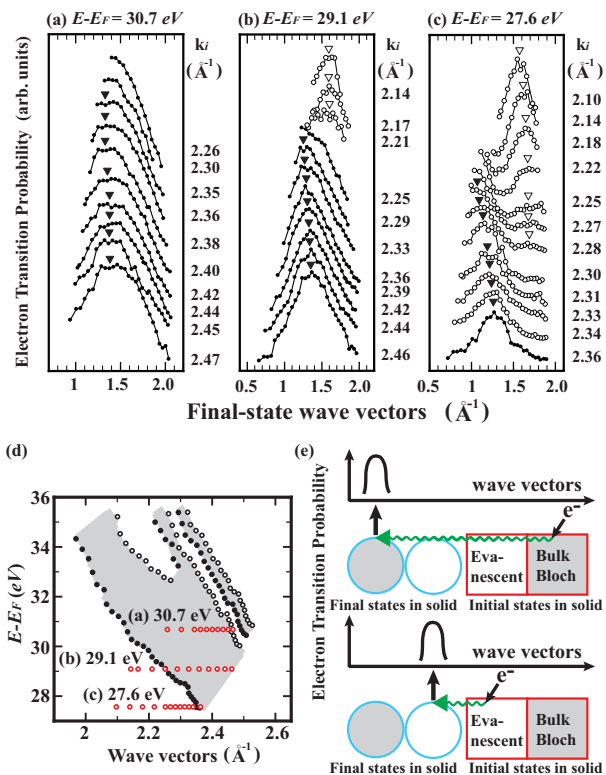}
\vspace{-10pt}
\caption{\label{fig:epsart}
(color online) (a)-(c) Electron-transition-probability spectra were taken for the initial states (a)-(c) in (d) (see the text). (d) A magnified VLEED band map with a group of initial states used in this work. The energy of each point in one set is the same. Each initial-state wave vector is shown at the right-hand side of each spectrum. The peaks are indicated by black and white triangles. (e) A scenario of the observed phenomena. The green arrows indicate one-phonon-excitation processes.}
\vspace{0pt}
\end{figure}


For each initial state, we measured the electron-transition-probability spectrum shown in Fig. 4(a)-(c). The spectra represented by black dots were taken of initial-state electrons that enter the bulk Bloch states located in the shaded area of the band map. On the other hand, the spectra represented by white-circles were taken of initial-state electrons that enter the evanescent states located in the white area of the band map. We observed a clear difference in the peak positions between the two spectra. Fig.~4(a) shows only a single peak in each spectrum, located at almost the same position, $K_{\parallel}^{S} \approx $1.3-1.4\AA$^{-1}$ and indicated by black triangles. In Fig.~4(b), one peak is found among the bottom spectra at almost the same positions, $K_{\parallel}^{S} \approx $1.3-1.4\AA$^{-1}$. However, in the spectra represented by white circles, another new peak, indicated by white triangles, appears at a different position, $K_{\parallel}^{S} \approx 1.6$\AA$^{-1}$. Fig.~4(c) shows the appearance of the new peak more clearly. The two peaks appear at distinctly different positions: $K_{\parallel}^{S} \approx $1.1-1.3\AA$^{-1}$ (indicated by black triangles) for the spectra represented by black dots and for some of the spectra represented by white circles~\cite{comment2} and $K_{\parallel}^{S} \approx$1.6-1.7\AA$^{-1}$ (indicated by white triangles) for the spectra represented by white circles. Note that the leftmost peak is present in most of the spectra in Fig. 4(c), and it shifts further to the left in the spectra represented by white circles highlighting the difference between Fig.~4(a) and 4(b).

The appearance of the new peak seen in Fig. 4(b) and (c) indicates a change in the specific final states that are optimally coupled to the initial states. This occurred because incident electrons enter different electronic states in solids, and the coupling between the initial and final states in the solid is changed, as illustrated in Fig. 4(e). Therefore, this result indicates that the measured electron transition probabilities reflect the strength of the coupling constants in elementary electron-phonon scattering processes in solids. By the same token, the drastic movement of the peak seen in Fig. 4(c) is also thought to be caused by changes in the final states. In this case, however, a clear change from one final state to the other was not observed in the form of two well-resolved peaks because of the limited resolution in the less well-measured region of the final-state wave vectors~\cite{comment2}.


Finally, we will comment on a possible application of the superconductivity of graphite. Although graphite is not an intrinsic superconductor, some graphite-intercalated compounds (GICs) such as C$_{6}$Ca and C$_{2}$Li exhibit superconductivity. The partial occupation of interlayer states drives superconductivity in these systems, but the microscopic features of superconductivity namely, coupling with phonons are still being explored~\cite{gic:1}. Although our method cannot address electronic states below the vacuum level E$_{vac}$, interlayer states can stay above E$_{vac}$ in some GICs, including graphite~\cite{vleed:2,gic:1}. If an initial state is set anywhere in the interlayer states, we expect to know which phonons optimally couple with the interlayer state by comparing the electron transition probabilities for phonons with different polarizations.

In summary, we have measured the modulation of electron transition probabilities in one-phonon-excitation processes in HREELS experiments using graphite as an example. Combining HREELS data with VLEED data, we have demonstrated that measured electron transition probabilities reflect the strength of the coupling constants in elementary electron-phonon scattering processes on a solid surface. This information could be applied to further explore the microscopic features of superconductivity in graphite intercalation compounds.

We acknowledge many useful discussions with Prof. K.-H. Rieder, Prof. M. Tsukada and Prof. V.N. Strocov. This work was supported in part by JSPS, a Grant-in-Aid for The 21st Century COE Program (Physics of Self-organization Systems) at Waseda University from the MEXT-Japan, and KAKENHI (20104008).



\begin{thebibliography}{99}


\bibitem{bcs:1}J. Bardeen {\it et al.}, Phys. Rev. {\bf 108}, 1175 (1957).
\bibitem{review1}V. Z. Kresin {\it et al.},  Rev. Mod. Phys. {\bf 81}, 481 (2009).

\bibitem{zimantext}J. M. Ziman, "Electrons and Phonons", Oxford University Press, New York, (2001). 

\bibitem{epi1:1}D. Pines, "Elementary excitation in solids", Benjamin, inc, New York, (1963). 


\bibitem{haugtext}H. Haug, and S. W. Koch, "Quantum theory of the optical and electronic properties of semiconductors", World Scientific, Singapore (2004). 


\bibitem{lisowski04}M. Lisowski{\it et al.}, Appl. Phys. A {\bf 78}, 165 (2004).

\bibitem{carbone09}F. Carbone {\it et al.}, Science {\bf 35}, 181 (2009).

\bibitem{epi:2}M. Hengsberger {\it et al.}, Phys. Rev. Lett. {\bf 83}, 592 (1999).
\bibitem{epi:3}P. G. Tomlinson {\it et al.}, Can. J. Phys. {\bf 55}, 751 (1977).

\bibitem{hreelsbook:1}H. Ibach, D. L. Mills, "Electron Energy Loss Spectroscopy and Surface Vibrations", Academic Press, New York, (1982).
\bibitem{hreels:1}H. Yanagisawa {\it et al.}, Phys. Rev. Lett. {\bf 93}, 177003 (2004).
\bibitem{hreels:2}H. Yanagisawa {\it et al.}, Surf. Inter. Anal. {\bf 37}, 133 (2005).

\bibitem{phononintensity:1}C. H. Li {\it et al.}, Phys. Rev. B {\bf 21}, 3057 (1980).

\bibitem{tong75}S. Y. Tong, Progress in surface science {\bf 7}, 1 (1975).

\bibitem{hufner95}S. Hufner, Photoelectron Spectroscopy: Principles and Applications (Springer-Werlag, Berlin, 1995).


\bibitem{vleed:4}N. Barrett {\it et al.}, Phys. Rev. B. {\bf 71}, 035427 (2005).

\bibitem{vleed:1}V. N. Strocov, Int. J. Mod. Phys. B {\bf 9}, 1755 (1995).


\bibitem{vleed:2}V. N. Strocov {\it et al.}, Phys. Rev. B {\bf 61}, 4994 (2000).



\bibitem{graphitefabrication:1}I. Forbeaux {\it et al.}, Phys. Rev. B {\bf 58}, 16396 (1998). 

\bibitem{graphitephonon:1}W. -H. Soe {\it et al.}, Phys. Rev. B {\bf 70}, 115421 (2004).

\bibitem{vleed:3}V. N. Strocov {\it et al.}, Phys. Rev. B {\bf 74}, 195125 (2006).


\bibitem{dacapo}The software package DACAPO can be downloaded at http://dcwww.camd.dtu.dk/campos/Dacapo/.

\bibitem{intensitymesurement:1}W. Ho {\it et al.}, Phys. Rev. B {\bf 21}, 4202 (1980).


\bibitem{comment2}It should be mentioned that the border between gap and bluk states can not be defined clearly like the border of band map in Fig.~2(c). Both kinds of electronic states must extend over the border to a certain extent, hence they are mixed in the vicinity of the border.


\bibitem{gic:1}G. Csanyi {\it et al.}, Nature phys. {\bf 1}, 42 (2005).

\end{thebibliography}
\end{document}